\documentclass[10pt,journal,compsoc]{IEEEtran}

\ifCLASSOPTIONcompsoc
  \usepackage[nocompress]{cite}
\else
  \usepackage{cite}
\fi
\ifCLASSINFOpdf
\else
\fi
\usepackage{url}            
\usepackage{tikz}
\usepackage{wrapfig}
\usepackage{graphicx}
\usepackage{xspace}
\usepackage{array}
\usepackage{booktabs}
\usepackage{multirow}
\usepackage{latexsym}
\usepackage[T1]{fontenc}
\usepackage{todonotes}
\usepackage{amsmath}
\usepackage{xcolor,colortbl}
\usepackage{hyperref}

\newcommand{\bc}{\textsc{BreastCancer}\xspace}
\newcommand{\pc}{\textsc{PancreaticCancer}\xspace}
\newcommand{\dm}{\textsc{DiabetesMellitus}\xspace}
\newcommand{\novelmi}{\textsc{Novel-miRNA}\xspace}
\newcommand{\noveld}{\textsc{Novel-Disease}\xspace}
\newcommand{\mi}{miRNA\xspace}
\newcommand{\midi}{miRNA-disease\xspace}
\newcommand{\approach}{\textsc{MuCoMiD}\xspace}
\newcommand{\mpara}[1]{\medskip\noindent{\bf #1}}
\newcommand{\ha}{\textsc{Hmdd2}\xspace}
\newcommand{\hb}{\textsc{Hmdd3}\xspace}
\newcommand{\epmda}{\textsc{Epmda}\xspace}
\newcommand{\dbmda}{\textsc{Dbmda}\xspace}
\newcommand{\nimgcn}{\textsc{Nimgcn}\xspace}

\newcommand{\dimig}{\textsc{DimiG 2.0}\xspace}
\newcommand{\nemii}{\textsc{NEMII}\xspace}
\newcommand{\ap}{\textsc{AP}\xspace}
\newcommand{\auc}{\textsc{AUC}\xspace}
\newcommand{\helda}{\textsc{Held-out1}\xspace}
\newcommand{\heldb}{\textsc{Held-out2}\xspace}
\newcommand{\minew}{\textsc{Novel-miRNA}\xspace}
\newcommand{\dinew}{\textsc{Novel-Disease}\xspace}

\begin{document}

\title{MuCoMiD: A \textbf{Mu}ltitask graph \textbf{Co}nvolutional Learning Framework for \textbf{mi}RNA-\textbf{D}isease Association Prediction}

\author{Ngan Dong$^{1}$, Stefanie Mücke$^{2}$, Megha Khosla$^{1}$\\
{\small $^{1}$ L3S Research Center, Leibniz University of Hannover, Hannover, Germany\\
$^{2}$ TRAIN Omics, Translational Alliance in Lower Saxony, Hannover, Germany\\
Email: dong@l3s.de}}

\markboth{}
{Shell \MakeLowercase{\textit{et al.}}: Bare Demo of IEEEtran.cls for Computer Society Journals}
\IEEEtitleabstractindextext{%

\begin{abstract}
Growing evidence from recent studies implies that microRNA or miRNA could serve as biomarkers in various complex human diseases. Since wet-lab experiments are expensive and time-consuming, computational techniques for miRNA-disease association prediction have attracted a lot of attention in recent years. Data scarcity is one of the major challenges in building reliable machine learning models. Data scarcity combined with the use of precalculated hand-crafted input features has led to problems of overfitting and data leakage. 

We overcome the limitations of existing works by proposing a novel multi-tasking graph convolution-based approach, which we refer to as \approach. MuCoMiD allows automatic feature extraction while incorporating knowledge from five heterogeneous biological information sources (interactions between miRNA/diseases and protein-coding genes (PCG), interactions between protein coding genes, miRNA family information, and disease ontology) in a multi-task setting which is a novel perspective and has not been studied before. 
To effectively test the generalization capability of our model, we construct large-scale experiments on standard benchmark datasets as well as our proposed larger independent test sets and case studies.  MuCoMiD shows an improvement of at least \textbf{$\sim 3$}\% in 5-fold CV evaluation on HMDDv2.0 and HMDDv3.0 datasets and at least \textbf{$\sim 35$}\% on larger independent test sets with unseen miRNA and diseases over state-of-the-art approaches. 
 We share our code for reproducibility and future research at~\href{https://git.l3s.uni-hannover.de/dong/cmtt}{\url{https://git.l3s.uni-hannover.de/dong/cmtt}}.
\end{abstract}
}

\maketitle
\IEEEdisplaynontitleabstractindextext

%
\IEEEpeerreviewmaketitle

\IEEEraisesectionheading{\section{Introduction}\label{sec:intro}}
Beginning in the early 2000s, the biological dogma that proteins are responsible for most functions in a cell began to shift and new classes of non-coding, regulatory RNAs became points of interest. The highly conserved class of microRNAs (miRNAs) with an approximate length of 22 nucleotides were first considered “junk” DNA without function and emerged as regulators in cell development, maturation, differentiation, and apoptosis of the cell, cell signaling, cellular interactions, and homeostasis~\cite{mattick2005small, kim2006genomics, saini2007genomic}. 

MicroRNAs fulfill their diverse functions by regulating gene expression of protein-coding genes (PCGs) after transcription. The transcribed mRNA of PCGs can be directly bound by miRNAs, which leads to cleavage or destabilization of the mRNA and represses the translation into proteins~\cite{cai2009brief}. Each miRNA can have hundreds of target mRNAs and mRNAs can be regulated by more than one miRNA, resulting in complex networks that are yet to be fully understood. The mutation of miRNAs or changes in their expression can have widespread consequences that can be hard to predict. 

Consequently several associations between miRNAs and diseases have been confirmed using biological experiments leading to the belief that miRNAs could be potential biomarkers in certain diseases such as cancers or immune-related diseases~\cite{de2018study, usuba2019circulating, jin2018serum, keller2011toward, schickel2008micrornas, zhang2007micrornas,lin2013characterization}. Besides unveiling a deeper understanding of diseases’ molecular pathogenesis, identifying potential associations between miRNAs and diseases can also help in the treatment and discovery of possible drug targets.

Owing to the significance of the problem and the time-consuming nature of biological experiments, recent years have seen an upsurge in machine learning approaches \cite{dong2019epmda,zheng2020dbmda,liu2020predicting,pan2020scoring,li2020neural} for \emph{predicting \midi associations}. 
Being successfully able to predict \midi associations can lead to information prioritisation in biological wet-lab experimentation, which will result in considerable time and cost savings.
From a machine learning perspective, the problem of predicting \midi associations can be formulated as that of a \emph{link prediction problem} in a bipartite graph, where \mi and disease form the two sets of nodes. 
However, as is typical to many biomedical applications, a major challenge for building generalizable and eventually well-performing models for the \midi association prediction problem is \emph{data scarcity} of \midi data. For example, the total number of \mi and disease nodes (without any preprocessing and removal of duplicates) in the standard HMDD v2.0 \cite{li2014hmdd} database are 495 and 383 respectively with a total of 5,430 associations (links). These networks are therefore not only small with respect to the nodes set but are also sparsely connected.

Data scarcity often leads to biased and non-generalizable models. 
Earlier attempts to address data scarcity rely on creating additional secondary features based on the initial feature set by computing intra-node similarities, e.g., \mi functional similarity.
So much so, that these secondary features that encode \mi functional similarity are pre-computed and are even deposited in well-known databases like MISIM~\cite{wang2010inferring}.
The first problem associated with using such similarity-based input features is that it is prone to overfitting because slight errors in the input features get amplified by the secondary features. 
Secondly, models based on secondary features cannot be used to predict new associations for a \mi (or a disease), i.e., instances for which no prior disease (or \mi) association information is available.

More worrying is the problem of \emph{data leakage} when using pre-computed \mi functional similarities indiscriminately from available databases. 
Dong and Khosla~\cite{dong2020consistent} find that most of the existing works that employ pre-computed similarities in model building ignore the actual train/test split giving rise to \emph{data leakage}.
In other words, some of the associations which are to be tested by the model are already present in the association network that was used to compute the similarity features.
Finally, the small dataset size for the \midi prediction task prohibits the utility of flexible and more expressive modern representation learning approaches.

\subsection{Present work}  We overcome the above limitations in the earlier literature by avoiding the creation of secondary features altogether.
Instead, our approach attempts to address the data scarcity problem by integrating knowledge from multiple heterogeneous sources of information available for \mi and diseases in addition to the \mi-disease associations. 
Combining multiple data sources enables us to compensate for missing or unreliable information in any single data type leading to more reliable predictions. 
Our key contribution is to model the integration of heterogeneous knowledge sources into a common representation space that can be trained end-to-end using modern representation learning machinery.

To this end, we propose a \textbf{Mu}ltitask graph \textbf{Co}nvolutational neural network for \textbf{mi}RNA-\textbf{D}isease association prediction, which we refer to as \approach for brevity. \approach allows automatic feature extraction while incorporating knowledge from five heterogeneous biological information sources (interactions between miRNA/diseases and protein-coding genes (PCG), interactions between protein-coding genes, miRNA family information, and disease ontology) in a \emph{multi-task setting}.  Instead of pre-calculated secondary features for miRNA and disease as in previous works, we employ graph convolution operation (without the non-linear activation) over the corresponding biological networks to learn informative representations for \mi and disease automatically at training time. Overall we have an expressive linear model with non-linear activation only at the output layer.

The added side tasks serve as regularizers and help us to incorporate domain-knowledge. For example, an miRNA $m$  regulates a set of proteins $p$ that are responsible for some biological functions. Moreover, disruptions in the biological functions of $p$ lead to certain disease condition $d$. Then $m$ has some influence over disease $d$ via $p$. The additional tasks of predicting \mi-PCG interaction and disease-PCG interaction help us to encode such influence by embedding $m$ and $d$ closer in the representational space. 
Besides, we employ \emph{adaptive loss balancing} techniques to fine-tune multi-task loss gradients. This allows us to utilize the full power of multi-task learning without resorting to exhaustive hyperparameter search. 

We conduct an extensive evaluation of our approaches in comparison to earlier works on existing benchmark datasets retrieved from HMDD v2.0~\cite{li2014hmdd} and HMDD v3.0~\cite{huang2019hmdd} databases.
In addition to standard benchmark datasets, we also construct and test on new and larger independent test sets. 
We finally present case studies for three specific diseases to showcase the utility of our approach in predicting the association of novel diseases for which no prior \mi association information is available in the train set.

\mpara{Our Contributions.} To summarize, we make the following contributions:
\begin{itemize}
\item We model the \midi association prediction problem in a \emph{multi-task setting} incorporating heterogeneous domain knowledge, which is a novel perspective and has not been studied before for the current problem. 

    \item We construct four new larger test sets for testing in transductive (when \mi and disease nodes in the test set are also present in the train set) and inductive settings (when the test set contains a number of new \mi or disease nodes). 
  
    \item We conduct large-scale experiments, ablation study and case studies for three specific diseases to showcase the superiority of our approach.  
    
    \item We release all the code and data used in this work for reproducibility and future research at~\href{https://git.l3s.uni-hannover.de/dong/cmtt}{\url{https://git.l3s.uni-hannover.de/dong/cmtt}}.
\end{itemize}

\section{Problem statement  and related work}
\label{secRelated}
The \midi association data can be represented using a bipartite graph $\mathcal{G}_{md} = (M,D,E)$ where $M$ is the set of nodes representing miRNAs and $D$ denotes the set of disease nodes. Each edge $e=(m,d)\in E$ denotes the association between the \mi node $m$ and disease node $d$.
We are then interested in the following problem statement.

{\itshape 
\mpara{Problem statement}
 Given the bipartite graph $\mathcal{G}_{md} = (M,D,E)$, we are interested in (i) predicting missing links among the given nodes (transductive setting) (ii) predicting new links for so far unseen novel \mi or disease nodes (inductive setting).
 }

\subsection{Related work}
 Existing computational approaches for \midi association prediction can be broadly grouped into three classes: \emph{scoring-based}, \emph{network topology} and \emph{machine learning} based methods. 
 
 Assuming that the \mi pairs linked to common diseases are functionally more related scoring based methods ~\cite{wang2010inferring,small2010micrornas} proposed scoring systems to prioritize \midi associations. A more sophisticated scoring scheme while integrating information from \mi and disease similarity networks was proposed in \cite{yang2017dbdemc}. Network based approaches~\cite{chen2012rwrmda,li2018predicting,chen2018heterogeneous} construct miRNAs and/or disease similarity networks and aims at efficiently transferring known miRNA-disease association labels between similar miRNAs and/or similar diseases in the network.  Chen et al.~\cite{chen2012rwrmda} employ repeated random walks with restarts over the \mi functional similarity network and prioritize candidate \midi associations using the final stable walk probability.

More closely related to our work is the third category of machine learning based methods. Approaches in this category  mainly rely on using secondary or hand-crafted features to construct similarity networks from which latent node features are extracted using graph-based representation learning techniques.
\epmda~\cite{dong2019epmda} extracts edge perturbation-based features from the \midi heterogeneous network and then trains a Multilayer Perceptron regression model to prioritize \midi associations. NNMDA~\cite{zeng2019prediction} combines information from five different miRNA similarities and two disease similarities to build a heterogeneous network for feature learning and association prediction. \cite{ji2020predicting} incorporates information from multiple domains, for example, miRNA-lncRNA and miRNA-PCG interaction, miRNA-drug association, disease-lncRNA, disease-PCG association, disease-drug association, to build a heterogeneous information network for feature extraction. The graph-based features along with \mi k-mer feature (calculated from the miRNA sequence) and disease semantic similarity are concatenated to form the input to a Random Forest classifier for association prediction. 

Another line of works include \nimgcn~\cite{li2020neural} and DimiG~\cite{pan2020scoring} which propose end-to-end learning approaches in which graph convolution networks (GCNs)\cite{kipf2017semi} are employed for extracting latent features of \mi and disease nodes. Non-graph based approaches like \dbmda~\cite{zheng2020dbmda} extract latent features from input hand-crafted features consisting of \mi functional, disease semantic, and \mi sequence similarity using autoencoders.
The latent features are then fed to a Rotation Forest~\cite{rodriguez2006rotation} classifier.

The reliance on hand-crafted features based on existing association data limits the applicability of existing techniques to transductive settings. Hence, apart from other limitations with respect to model generalization and data leakage (already discussed in Section \ref{sec:intro}) a majority of these approaches cannot be applied to predict associations for new \mi or disease nodes that have not been observed in training data.

\section{Proposed approach}
\label{secModel}
\begin{figure*}[ht!]
    \centering
        \includegraphics[width=0.9\linewidth]{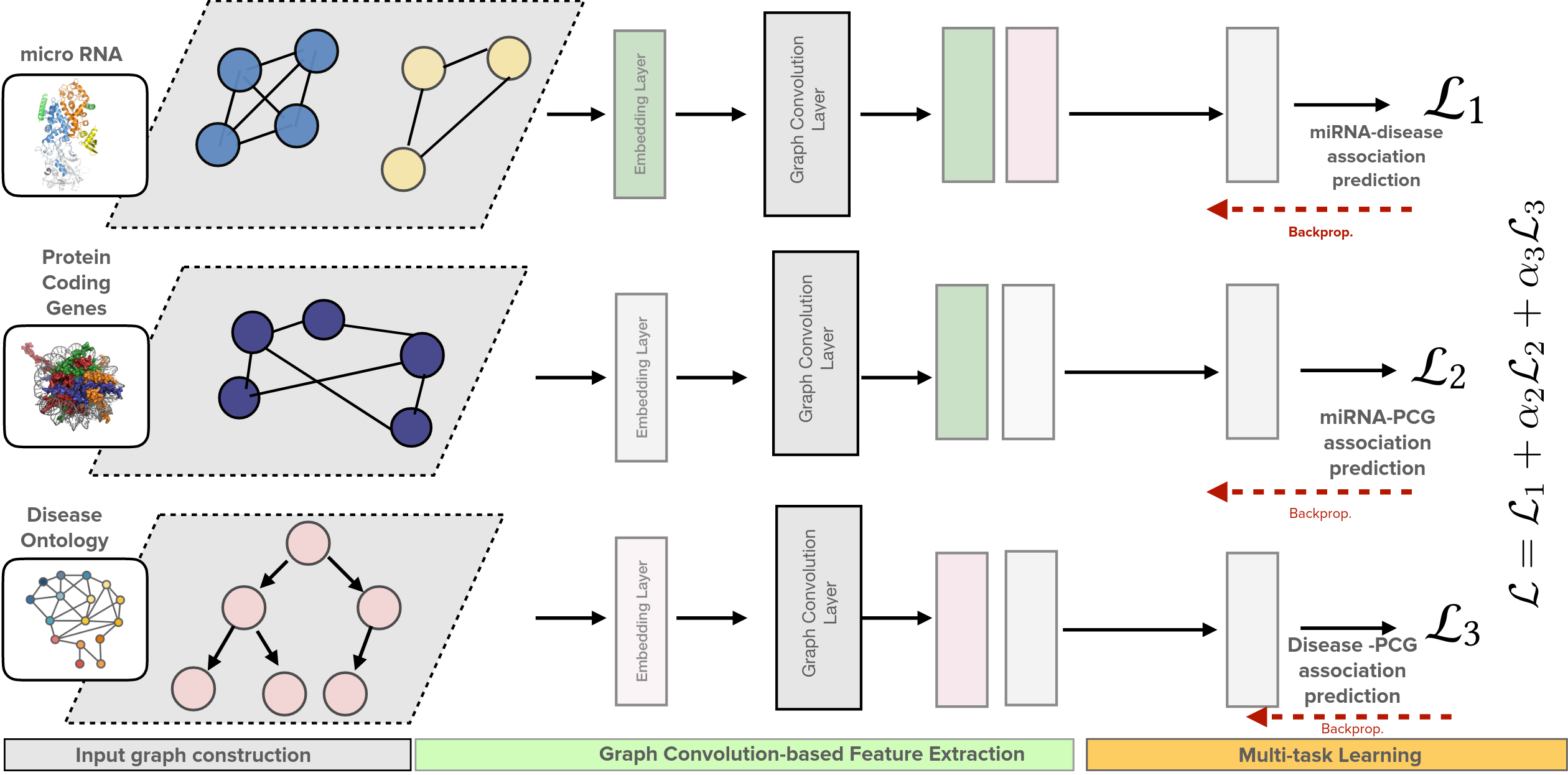}
    \caption{A schematic diagram of \approach. \approach consists of three main modules: (i) input graph construction in which we build networks corresponding to the available side information from \mi family, PCG-PCG interactions, and disease ontology (ii) the second module takes the constructed networks as input and generates the nodes' representation according to their local neighbors (iii) finally, one classifier for \midi association prediction, two regressors for miRNA-PCG and disease-PCG interaction confidence score prediction are added. The second and third modules get trained jointly using a multitask loss. The multitask loss is a weighted sum of the three individual task losses and is optimized using a \emph{dynamic loss balancing} technique.}
    \label{fig:network}
\end{figure*}
Given the input bipartite \midi network $\mathcal{G}_{md}$, we treat the \midi association problem as a binary classification where the label for an input pair node $(m,d)$ is $1$ if there is a known association between them and $0$ otherwise.

To overcome the challenges of data scarcity, we propose \approach in which we focus on \emph{effectively integrating heterogeneous biological information while learning to predict missing or new \midi associations}. Instead of relying on secondary or handcrafted features, we rely on other knowledge sources to learn the input representation automatically at training time. Also recent works indicate that PCGs are the most important link between miRNAs and their associated diseases~\cite{mork2014protein} since changes in miRNAs lead to differently regulated PCGs, which in turn can cause diseases. Therefore, in addition to the \midi association prediction task (formulated as a binary classification problem), we add miRNA-PCG and disease-PCG interaction confidence score prediction (formulated as regression tasks) as two additional side tasks to incorporate additional domain knowledge and prevent overfitting. 

In summary, \approach employs different ways of integrating domain knowledge at different stages of the model building process. In particular, information from three different biological networks namely, miRNA family , PCG-PCG interaction, and disease ontology is directly used as inputs to learn the node representations during training. Besides, the miRNA-PCG and disease-PCG interactions are employed to build additional regularization objective functions. The incorporation of various information sources help compensate for the lack of information in single data source. It also helps in mitigating the data scarcity problem.

In the following we describe the three modules of \approach (also depicted in Figure \ref{fig:network}): (i) \emph{input graph construction}, (ii) \emph{graph convolution based feature extraction}, and (iii) \emph{multi-task optimization/learning}.

\subsection{Input graph construction}
\label{sec:graphs}
We start by describing the construction or retrieval of various biological networks that we leverage as additional sources of information and the corresponding rationale.

\begin{figure}[h!]
    \centering
    \includegraphics[width=0.8\linewidth]{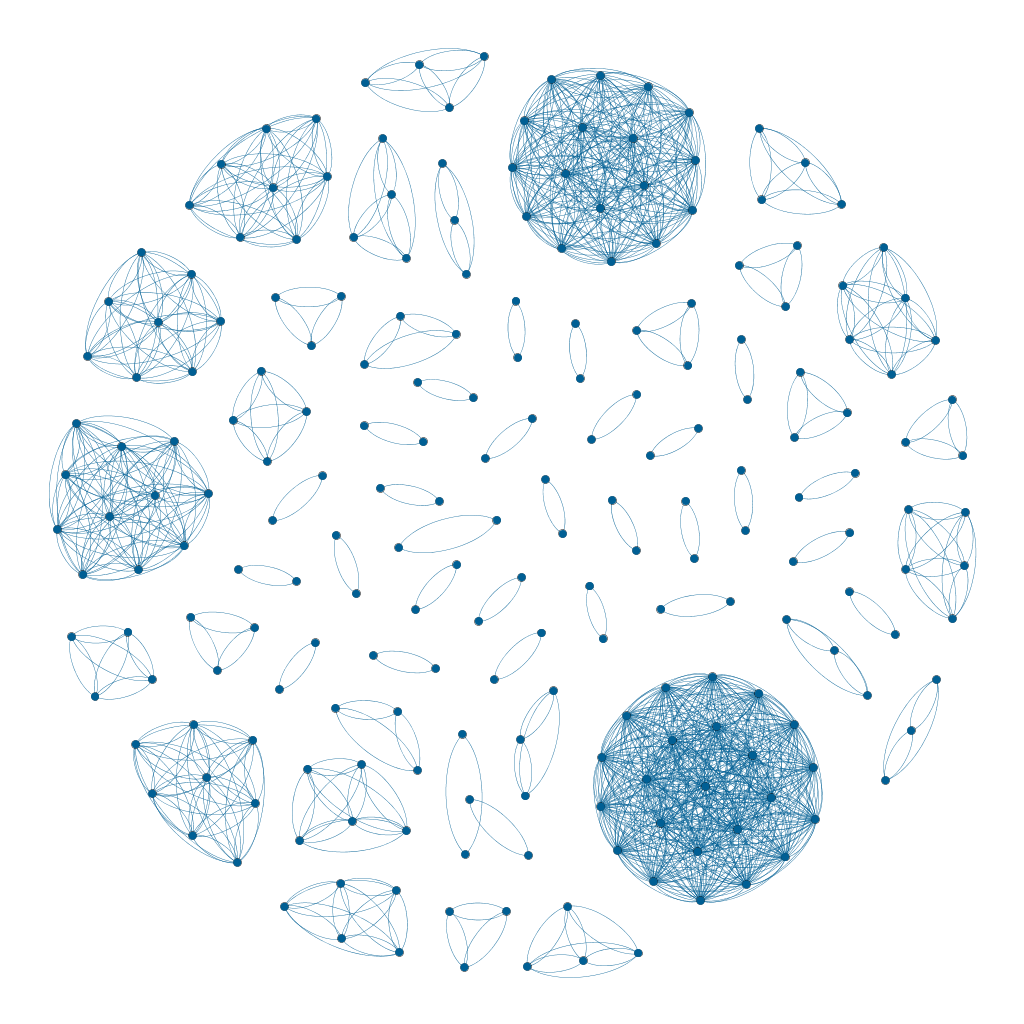}
    \caption{An illustration of the miRNA family network.}
    \label{fig:miFam}
\end{figure}

\mpara{\mi family, $\mathcal{G}_m$}. A miRNA family is the group of miRNAs that share a common ancestor in the phylogenetic tree. miRNAs that belong to the same family usually have highly similar sequences, secondary structures and tend to execute similar biological functions~\cite{kaczkowski2009structural}. Similar \mi would tend to participate in the mechanisms of similar diseases. We retrieve \mi family information from mirBase database~\cite{kozomara2010mirbase}. The \mi network $\mathcal{G}_m$ is an unweighted undirected graph in which there is a connection between node A and node B if A and B belong to the same family. Figure~\ref{fig:miFam} presents an illustration of the miRNA family network generated from our data.

\begin{figure}[h!]
    \centering
    \includegraphics[width=0.7\linewidth]{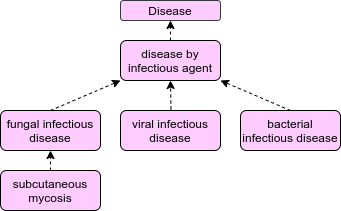}
    \caption{An illustration of the disease ontology. Directed edges denote `\textit{is\_a}' relations}
    \label{fig:diseaseOnto}
\end{figure}

\mpara{Disease ontology, $\mathcal{G}_d$}. The disease ontology~\cite{schriml2012disease} represents the disease etiology classes. A directed connection between two diseases exists if there exists a \textbf{is-a} relationship between them. Similar diseases can be expected to interact with similar miRNAs. The disease ontology network $\mathcal{G}_d$ is an unweighted directed network in which there is a directed connection from A to B if B is a parent of A. $\mathcal{G}_d$ can be visualized as a directed tree which contains only directed connection between children and parents nodes. Each tree layer represents one layer of abstraction. The upper most layer represents the most general disease category. An illustration of the disease ontology is given in figure~\ref{fig:diseaseOnto}.

\begin{figure}[h!]
    \centering
    \includegraphics[width=0.7\linewidth]{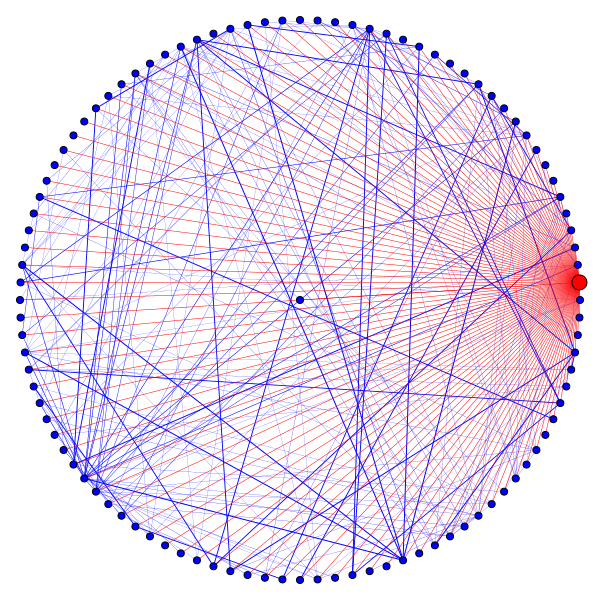}
    \caption{An illustration of miRNA-PCG, PCG-PCG interactions. Red nodes and red edges denote miRNAs and miRNA-PCG connections while blue nodes and blue edges represent PCGs and PCG-PCG interactions, respectively.}
    \label{fig:mipcg}
\end{figure}

\mpara{PCG-PCG interaction, $\mathcal{G}_p$.} PCGs interact with PCGs to carry out biological functions.  Therefore, given the fact that protein coding gene $p_1$ activates the expression of protein coding gene $p_2$, if the \mi $m$ can regulate $p_1$  then there should be some relation between $m$ and $p_2$. In other words, information from the protein-protein interaction network will bring additional insights about indirect relationship between miRNAs/diseases and the rest of PCGs with which a direct interaction in not known.  An example of the \mi-PCG, PCG-PCG interactions is presented in figure~\ref{fig:mipcg} where red nodes and red edges denote miRNAs and miRNA-PCG connections while blue nodes and blue edges represent PCGs and PCG-PCG interactions, respectively. We download PCG interaction data from STRING v10 database~\cite{szklarczyk2015string}. As preprocessing step, we retain only interactions between PCGs that have at least one known interaction with miRNAs or diseases. The PCG network $\mathcal{G}_p$ is an undirected weighted network in which the edge weights correspond to the  confidence score of PCG-PCG interaction retrieved from STRING database.
\subsection{Graph convolution-based feature extraction} 
\label{secRep}

Having constructed the relevant networks we next extract informative node representations using the node neighborhood information. As we have no input node features, we train embedding layers (of dimension 32) for each of the network.

An embedding layer is essentially a look-up table where $i$th row corresponds to the learnt representation of $i$th node. 
The node embedding is then passed as an input feature to the graph convolutional layer. A graph convolutional layer is essentially a linear layer which transforms the node feature as an aggregation of representations of its 1-hop neighbors. In particular for input adjacency matrix $\mathbf{A}$ and node embedding matrix $\mathbf{X}$ we obtain transformed node feature matrix $\mathbf{X}'$ as follows
\begin{align}
\label{eqn:gcn}
  \mathbf{X}{'} = \mathbf{\hat{D}}^{-1/2} \hat{\mathbf{A}}
        \hat{\mathbf{D}}^{-1/2} \mathbf{X} \mathbf{W},
\end{align}
where $\hat{\mathbf{A}}= \mathbf{A} + \mathbf{I}$, $\mathbf{I}$ is the identity matrix and $\hat{\mathbf{D}}$ is the degree matrix of $\hat{\mathbf{A}}$ and $\mathbf{W}$ is trainable weight matrix of the graph convolutional layer.
We point out that we don't use any non -linear activations in this module, and the aggregation operation in Equation \eqref{eqn:gcn} can be efficiently carried out using sparse matrix multiplications. Moreover, we avoid the issue of non-interpretability in the feature extraction module, which is usually associated with non-linearities.

As the graph semantics are different for each of the networks, no parameter sharing is employed at this stage. We use three separate graph convolutional layers to extract representations for \mi, PCG and disease nodes. These learned representation will be fed as input to the multi-task optimization/learning module explained in the next section. 

\subsection{Multi-task optimization/learning}
To effectively utilize information from \mi-PCG and disease-PCG interactions, we design a multitask objective to train our model. In particular, for \mi-disease, \mi-PCG, and disease-PCG pairs, we construct association representations by taking an elementwise product of the corresponding node features. For example, for an \mi-disease input pair $(m,d)$ denoted by nodes $m$ and $d$ respectively, we obtain the corresponding feature vector representation as
$$\mathbf{x}_{md}^{\prime} = \mathbf{X}^{\prime}_m \odot {\mathbf{X}}^{\prime}_m$$
where $\mathbf{X}^{\prime}_m$ and $\mathbf{X}^{\prime}_d$ correspond to output representations of the corresponding graph convolution layer for nodes $m$ and $d$ respectively.

Using the pairwise representations, we then predict the existence of an association between \mi and disease; and the confidence score of associations for \mi-PCG and disease-PCG pairs. In summary, we train our model with multi-task loss functions calculated from these three supervised tasks and use an adaptive loss balancing technique to combine the three individual loss components dynamically at training time. Details about individual task loss and our optimization strategy are presented in the following sections.

\subsubsection{\midi binary classification task loss ($\mathcal{L}_{1}$).} We compute the probability of observing an association between an \midi input pair $(m,d)$ as:
\begin{align}
\label{eq:mdredict}
    y_{md}= \sigma\left(\mathbf{w}^{T}_{MD} \mathbf{x}'_{md}\right)
\end{align}
where $\mathbf{w}_{MD}$ is a learnable weight matrix and $\sigma(x)= {1\over 1+ \exp(-x)}$ is the sigmoid function. We use binary cross entropy to calculate the training loss for the \midi classification module as follows:
\begin{align}
\label{eq:lossmd}
    \mathcal{L}_{1}= \sum_{m,d} -z_{md} \log y_{md} -(1-z_{md}) \log(1-y_{md})
\end{align}
where $z_{md}$ denote the target label known for the corresponding training pair.
\subsubsection{miRNA-PCG regression task loss ($\mathcal{L}_{2}$).} For an input miRNA-PCG pair $(m,p)$, we compute the association confidence score as:
\begin{align}
\label{eq:mppred}
    y_{mp}= \sigma\left(\mathbf{w}^{T}_{MP} \mathbf{x}'_{mp}\right)
\end{align}
where $\mathbf{w}_{MP}$ is a learnable weight matrix and $\sigma(x)$ is the sigmoid function. We use the sum of squared error to calculate the training loss for the miRNA-PCG regression module as follows:
\begin{align}
\label{eq:l2}
    \mathcal{L}_2 = \sum_{m,p} (y_{mp} - z_{mp})^{2}, 
\end{align}
where $z_{mp}$ denotes the target confidence score.

\subsubsection{disease-PCG regression task loss ($\mathcal{L}_{3}$).}
Similar to prediction of \mi-PCG association, we compute for a disease-PCG input pair $(d,p)$ the association confidence score, $y_{dp}$. $\mathcal{L}_{3}$ is then calculated using the sum of squared error as in~\ref{eq:l2}. 
\begin{align}
\label{eq:l3}
    \mathcal{L}_3 = \sum_{d,p} (y_{dp} - z_{dp})^{2}, 
\end{align}
where $z_{dp}$ denotes the target confidence score.

\subsubsection{Multi-Task optimization}
We define the final loss for our model as the linear combination of three losses as follows:
\begin{align}
\label{eq:mtt}
    \mathcal{L}= \mathcal{L}_1 +  \alpha_2 \mathcal{L}_2 +  \alpha_3 \mathcal{L}_3,
\end{align} 
where $\alpha_2$, and $\alpha_3$ are the loss weight for the two side tasks. Generally, multi-task networks are difficult to train. Finding the optimal combination of individual task losses is challenging and problem-specific. A task that is too dominant during training will overwhelm the update signals and prevent the network parameters from converging to robust shared features that are useful across all tasks.

We update $\alpha_2$, and $\alpha_3$ so that the difference between the two side tasks contribution at each time step $t$ is minimized. More specifically, at each time step $t$, the values for $\alpha_2$, and $\alpha_3$ are computed dynamically as follows:
\begin{align}
    \alpha_2(t) = \frac{\mathcal{L}_3(t-1) }{\mathcal{L}_1(t-1) + \mathcal{L}_2(t-1) + \mathcal{L}_3(t-1) + 10^{-1}}\\
    \alpha_3(t) = \frac{\mathcal{L}_2(t-1) }{\mathcal{L}_1(t-1) + \mathcal{L}_2(t-1) + \mathcal{L}_3(t-1) + 10^{-1}}
\end{align}

We use Adam optimizer with a learning rate of $10^{-3}$ to train the multi-task model.

\section{Experimental setup}
\label{secSetup}
\subsection{\mi-disease association datasets}
\label{sec:hmdd23}
We retrieve the set of \midi associations from the HMDD v2.0 database~\cite{li2014hmdd} and HMDD v3.0 database~\cite{huang2019hmdd}. As pre-processing steps, we retain only associations for miRNAs and diseases for which the PCG interaction information is available. The filtered data for the HMDD v2.0 database, which from now on is denoted as \ha, contains 2,303 known associations between 368 \mi and 124 diseases. The filtered data for the HMDD v3.0 database, which from now on is referred to as \hb, includes 8,747 known associations between 710 miRNAs and 311 diseases. 

\label{secData}

\begin{table}[!htb]
    \centering
    \caption{The data statistics where $|E|$, $|V_{miRNA}|$, $|V_{disease}|$ refer to the number of associations/links, miRNAs and diseases respectively.}
    \begin{tabular}{lccc}
         \textsc{Dataset} & $|E|$ & $|V_{miRNA}|$ & $|V_{disease}|$   \\
         \toprule
         \ha & 2,303 & 368 & 124  \\
         \hb & 8,747 & 710 & 311  \\
    \end{tabular}
    \label{tab:data_stats}
\end{table}

\subsection{miRNA-PCG interaction.}
\label{sec:mirnapcg}
We obtain the \mi-PCG interactions from the RAIN database~\cite{junge2017rain}. We include only interactions for the PCGs with at least one associated Reactome pathway~\cite{fabregat2018reactome} as these would be biologically more significant. We use all known \mi-PCG pairs to train the mi-PCG interaction confidence score prediction task and the target variables are set to be the normalized confidence score of the interaction retrieved from the database.

\subsection{disease-PCG interaction.}
\label{sec:diseasepcg}
We obtain the disease-PCG associations from the DISEASES database~\cite{pletscher2015diseases}. Here also, we retain only the associations for the PCGs with at least one associated Reactome pathway. As above, we use all known disease-PCG pairs to train the disease-PCG interaction confidence score prediction side task and the target variables are set to be the normalized confidence score of the known associations retrieved from the database.

Table \ref{tab:data_side} provides statistics of the three biological networks as described in section~\ref{sec:graphs} and the two additional datasets described in section~\ref{sec:mirnapcg} and~\ref{sec:diseasepcg}.

\begin{table}[!htb]
    
    \caption{Statistics for datasets with side information. $|E|$ is the number of connections/associations. $|V_{m}|$, $|V_{d}|$, and $|V_{P}|$ are the number of miRNAs, diseases, and PCGs, correspondingly.}
    \begin{tabular}{lrccc}
         \textsc{Network} & $|E|$ & $|V_{m}|$ & $|V_{d}|$ & $|V_{P}|$ \\
         \toprule
         \textsc{miRNA-PCG} & 178,716 & 714 & - & 9,236 \\
         \textsc{disease-PCG} & 144,846 & - & 312 & 9,236 \\
         \textsc{\mi family} ($\mathcal{G}_{m}$) & 1,354 & 217 & - & -\\
         \textsc{disease ontology} ($\mathcal{G}_{d}$) & 90 & - & 128 & -\\
         \textsc{PCG-PCG} ($\mathcal{G}_{p}$)& 4,446,583 & - & - & 9,236 \\
    \end{tabular}
    \label{tab:data_side}
\end{table}

\subsection{Our new test sets}
\label{sec:newtest}
 For small-size datasets like \ha and \hb, 5-fold CV evaluation is limited as the size of the training and testing sets become much smaller. While one can use \ha for training and \hb for testing, such evaluation is limited as there are many overlapping associations in these two datasets. We, therefore, carefully construct the following four independent tests using the \hb dataset. \textit{\ha is used as the training set for evaluation with the new test sets}. Let $\mathbf{M2}$ be the set of all miRNA in \ha and $\mathbf{D2}$ be the set of all diseases in \ha. The construction of the four independent test sets is described below.

\begin{table}[!htb]
    \centering
    \caption{Statistics for our new test datasets.}
    \begin{tabular}{lccc}
         \textsc{Dataset} & $|E|$ & $|V_{miRNA}|$ & $|V_{disease}|$   \\
         \toprule
         \helda & 2,669 & 324 & 110  \\
         \heldb & 6,641 & 692 & 303 \\
         \minew & 3,575 & 577 & 115  \\
         \dinew & 5,308 & 346 & 295  \\
    \end{tabular}
    \label{tab:testdata_stats}
\end{table}

\mpara{\helda for transductive testing.} To construct our \helda test set we include all \mi in $\mathbf{M2}$ and disease nodes in $\mathbf{D2}$ which are common in \ha and \hb. The \helda test set contains only the associations which are present in \hb but not in \ha. This is a transductive setting for link/association prediction such that all nodes in the test set are the same as that in the training set. We randomly generate the same number of negative samples from the set of unknown \midi pairs. Finally, \helda contains 2,669 known associations between 324 miRNAs and 110 diseases.

\mpara{\heldb for inductive testing.} 
We construct the \heldb test set by including all \mi and disease nodes and their known associations that are present in \hb but not in \ha. Note that different from \helda, \heldb might also contain associations corresponding to \mi and disease nodes which are not present in the training set \ha.
We randomly generate the same number of negative samples from the set of unknown \midi pairs. \heldb consists of 6,641 known associations between 692 miRNAs and 303 diseases.

\mpara{\textsc{Novel-miRNA}.}
We first fix the set of diseases in our test set to those retained from \ha, i.e., the set $\mathbf{D2}$ . Then all known associations that are (i) between diseases in $\mathbf{D2}$ and any miRNA and (ii) present in \hb but not in \ha are included in the \textsc{Novel-miRNA} test set.

We randomly generate the same number of negative samples from the set of unknown \midi pairs. In the end, our unseen \mi testing set consists of 3,575 associations between 577 miRNAs and 115 diseases. We use this set to test the performance of models on unseen data where there are new miRNAs for which no prior association is present in the training set.

\mpara{\noveld.}
Similarly, for constructing the \noveld test set, we first restrict the set of known associations to only those that involve miRNAs in $\mathbf{M2}$. Then we retain only associations that are in \hb but not in \ha.
We randomly generate the same number of negative samples from the set of unknown \midi pairs. Finally, we attain 5,308 associations between 346 miRNAs and 295 diseases. We use this set to test the performance of models on unseen data where there are new diseases for which no prior association is present in the training set.

\begin{figure}
    \centering
    \includegraphics[width=0.8\linewidth]{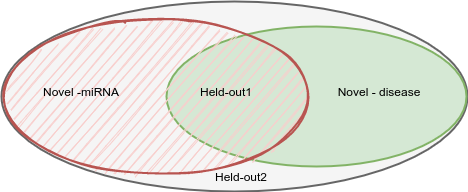}
    \caption{An illustration of the four large independent test sets relations where \helda = \novelmi $\cap$ \noveld, \heldb = \novelmi $\cup$ \noveld = \hb \textbackslash $ $ \ha.}
    \label{fig:data}
\end{figure}

A schematic Venn diagram of the four large independent test sets is presented in Figure~\ref{fig:data}.

\subsection{Benchmarked models}
We compare our model with five recently proposed supervised methods. Details about our benchmarked models are given below. 

\mpara{\epmda~\cite{dong2019epmda}}
\epmda consists of two components: the feature extractor and the classifier. The feature extraction module adopts a network topology-based approach and operates on a heterogeneous network $G$. $G$ is constructed from the known \midi association set, the pre-calculated \mi, and disease Gaussian Interaction Profile Kernel (GIP)~\cite{van2011gaussian} similarities. A Multiple Layer Perceptron (MLP) classifier is employed to do the classification task. \epmda feature extractor and classifier get trained separately.

\mpara{\textsc{\nemii~\cite{gong2019network}}}
\nemii uses Structural Deep Network Embedding (SDNE) to learn the embedding for \mi and disease from a bipartite graph built up from known association information. The learned embedding and features extracted from the \mi family and disease semantic similarity are concatenated to form the input to a Random Forest classifier. 

\mpara{\nimgcn~\cite{li2020neural}}
\nimgcn proposes an end-to-end learning framework that operates on a heterogeneous network $G$, which is built up from \midi known associations, the pre-calculated \mi functional similarity (MISIM), and the disease semantic similarities. GCN and non-linear transformation layers are employed to learn the latent representation for miRNAs and diseases separately. For a particular \midi pair, its association probability is calculated as the inner product of the two corresponding latent feature vectors.

\mpara{\dbmda~\cite{zheng2020dbmda}.}
\dbmda separates the feature learning and classification process. It uses autoencoders to learn the hidden representation for each \midi pair from pre-calculated similarities as input. The autoencoders are trained in an unsupervised manner. The encoded representation will then be fed as input into a Rotation Forest classifier whose job is to predict potential \midi associations.

\mpara{\textsc{\dimig~\cite{pan2020scoring}}}
\dimig is a semi-supervised approach that treats \midi association prediction as a multiclass classification problem where diseases are the labels. They do not use \midi association during the training process. Instead, they use only the known disease-PCGs interaction to learn the model parameters. PCGs nodes are connected with miRNAs nodes in a heterogeneous network based on miRNAs interaction profiles. Learned signals are then propagated through the heterogeneous network to infer the labels for miRNAs.

Note that from all the above baselines, only \dimig can be used for testing in inductive settings, i.e., testing for associations/links when the corresponding nodes are not present in the train set. 

\subsection{Test set-up and evaluation}
As in previous works, we perform 5-fold CV for testing on \ha and \hb datasets. We run 5-fold CV with 5 random initializations. In other words, for each dataset, we run each model $5\times 5=25$ times and report the average performance with the standard deviation.

To test on our new test sets, we train the models on the \ha dataset. We run the experiments 5 times with random initializations and report the mean performance scores along with standard deviation. For case studies, we run each model with the given case study data 5 times and report average performance for each model.

\mpara{Evaluation metrics.} Following previous works and recommendations in \cite{dong2020consistent}, we report the Area under the Receiver Operating Characteristic (\auc), and Average Precision(\ap) as our evaluation criteria. For our case studies, we report the number of known associations found in the top K pairs with the highest prediction scores returned from the model, where $k$ is from 10 to 100 with a step of 10. 

\subsection{Hyperparameter settings}
\subsubsection{\approach}
In all experiments, we fix the number of training epochs to 200, the embedding size and the hidden dimension both to 32. We employ Adam optimizer with a learning rate of $10^{-3}$ for training.

\subsubsection{Benchmarked models}
For \epmda, \dbmda and \nimgcn, we use the code and setup released in~\cite{dong2020consistent}. For \textsc{\nemii}, we use the same code and setup as published by the authors. 

For \textsc{\dimig}, we use the code and parameters shared by the author. To test the model performance on our data, we compared \textsc{\dimig} with the input features as tissue expression profiles and \textsc{\dimig} with one-hot vectors on the subset of our testing datasets that have \mi expression profiles available. The two models acquire similar performance. This implies that use of tissue expression profiles as node features do not have an effect on the model performance. We can therefore the test the model on data for which tissue expression information is unavailable by using one hot encoding for input node features. The results reported in Section~\ref{secResult} corresponds to the model with one-hot vectors as input features on our testing datasets without removing miRNAs that don't have expression profiles.

\section{Results}
\label{secResult}
 In the following, we discuss the performance gains of our approach (and their implications) on a variety of testing scenarios and new test datasets described in sections \ref{sec:hmdd23} and \ref{sec:newtest}.

\subsection{Results on small test sets}

\begin{table*}[ht!]
    \centering
     \caption{Results with 5FoldCV over the \ha and \hb  datasets. We perform 5-Fold CV 5 times with different random seeds. We report the average over 25 scores along with standard deviation. Our improvements over state-of-the-art (SOTA) methods are statistically significant with a p-value less than 0.002 on \ha and less than $10^{-30}$ on \hb datasets.}
  \begin{tabular}{l|c|c|c|c}
    Method & \multicolumn{2}{c}{\textsc{HMDD2}} & \multicolumn{2}{c}{\textsc{HMDD3}}\\
    & AUC & AP & AUC & AP\\
    \toprule
    \epmda (\cite{dong2019epmda}) & 0.744 $\pm$ 0.019 & 0.783 $\pm$ .017 & 0.520 $\pm$ 0.011 & 0.594 $\pm$ 0.010 \\
\nimgcn (\cite{li2020neural})& 0.785 $\pm$ 0.018 & 0.803 $\pm$ 0.015 & 0.795 $\pm$ 0.021 & 0.800 $\pm$ 0.018 \\
\dbmda (\cite{zheng2020dbmda}) & 0.553 $\pm$ 0.019 & 0.537 $\pm$ 0.015 & 0.749 $\pm$ 0.010 & 0.696 $\pm$ 0.009 \\
\nemii (\cite{gong2019network}) & 0.499 $\pm$ 0.019 & 0.502 $\pm$ 0.020 & 0.512 $\pm$ 0.008  & 0.509 $\pm$ 0.006 \\
\dimig (\cite{pan2020scoring})& 0.493 $\pm$ 0.018 & 0.485 $\pm$ 0.012 & 0.516 $\pm$ 0.006 & 0.508 $\pm$ 0.005 \\
\midrule
\textsc{\approach(ours)} & \bf{0.833} $\pm$ 0.012 & \bf{0.832} $\pm$ 0.015 & \bf{0.915}  $\pm$ 0.004 & \bf{0.910} $\pm$ 0.006 \\
\midrule
Improvement over SOTA & 6.1\% & 3.6\% & 15.1\% & 13.8\% \\
    \end{tabular}
   \label{tab:5FoldCV} 
\end{table*}

Following previous works, we perform 5-fold CV experiments on \ha and \hb datasets. The results are shown in Table \ref{tab:5FoldCV}.
The test set size for this scenario is considerably smaller and contains only  $1/5$th of the total associations. The test size is then considerably smaller for datasets like \ha.
Such a train-test scenario allows us to quantify how well the models learn but is limited in testing the generalization power of the models.

 While the graph-based approaches \epmda and \nimgcn perform reasonably well in this scenario, our multi-task-based approaches significantly supersede state-of-the-art (SOTA) methods with an improvement of up to $\sim$13.8\% in \ap score.
 
The performance of \epmda drops considerably for \hb dataset. \epmda learns edge features in an unsupervised manner corresponding to its contribution to a cycle of a particular length. Usually, the cycle length parameter is fixed to a small value due to an exponential increase in run time with an increase in cycle length. Moreover, the task signal is not used in learning the edge features. The loss of performance of \epmda in \hb can be attributed to the limitation of finding the best cycle length hyperparameter applicable for \hb. This also limits the applicability of this model to a larger variety of datasets. \nimgcn performs better than \dbmda due to the higher representational capacity of used GCNs and exploitation of additional graph structure.

\subsection{Results on small train but larger test sets in transductive setting}
\begin{table}[!ht]
    \centering
    \caption{Results on the \helda dataset after five runs. Our improvements over SOTA methods are statistically significant with a p-value less than $10^{-15}$.}
    \begin{tabular}{l|c|c}
    Method& AUC & AP\\
    \toprule
    \epmda & 0.562 $\pm$ 0.085 & 0.511 $\pm$ 0.064 \\
\nimgcn & 0.676 $\pm$ 0.009 & 0.642 $\pm$ 0.006\\
\dbmda  & 0.545 $\pm$ 0.011 & 0.521 $\pm$ 0.009\\
\nemii & 0.504 $\pm$ 0.002 & 0.501 $\pm$ 0.000 \\
\dimig & 0.429 $\pm$ 0.002 & 0.471 $\pm$ 0.003 \\
\midrule
\approach (ours) & 0.705 $\pm$ 0.005 & 0.691 $\pm$ 0.006 \\
Improvement over SOTA & 4.3\% & 7.6\% \\
    \end{tabular}
    \label{tab:heldout1}
\end{table}
Table \ref{tab:heldout1} shows results corresponding to \helda test set. Recall that \ha is used as the training set. In this scenario, the test size is much larger than the train set size allowing us to compare the generalization capability of the models.  

Out of the state-of-the-art models, \nimgcn is the best owing to the use of GCNs to extract latent representations. The overall drop in performance of all models in this scenario as compared to small test size case points to the hardness of this particular test set. Existing methods show a bad generalization capability.

Though both \approach and \nimgcn employ GCNs to extract latent representations, we still observe a significant difference between the two models' performance. The reason for such difference are: (i) \nimgcn is a single task-based model which relies on pre-calculated secondary features and learns solely from the set of known associations while (ii) \approach is a multitask-based model which can incorporate information from five additional knowledge sources to learn the feature automatically at run time and to overcome the limited training data problem. 

\subsection{Results on small train but larger test sets in inductive setting}
\begin{table*}[!ht]
    \centering
    \caption{Results on large test sets with new \mi and disease after five runs. Our improvements over SOTA methods are statistically significant with a p-value less than $10^{-15}$. All models are trained on \ha dataset.}
    \begin{tabular}{l|c|c|c|c|c|c}
    Method & \multicolumn{2}{c}{\minew} & \multicolumn{2}{c}{\dinew}& \multicolumn{2}{c}{\heldb}\\
    & AUC & AP & AUC & AP & AUC & AP \\
    \toprule
\dimig & 0.452 $\pm$ 0.001 & 0.480 $\pm$ 0.001 & 0.421 $\pm$ 0.001 & 0.467 $\pm$ 0.001 & 0.417 $\pm$ 0.003 & 0.465 $\pm$ 0.004 \\
\approach & \bf{0.726} $\pm$ 0.005 & \bf{0.713} $\pm$ 0.006 & \bf{0.637} $\pm$ 0.019 & \bf{0.632} $\pm$ 0.025 & \bf{0.648} $\pm$ 0.019 & \bf{0.662}  $\pm$ 0.025\\
\midrule
Improvement over SOTA & 60.6\% & 48.5\% & 51.3\% & 35.3\% & 55.4\% & 42.4\%\\
    \end{tabular}
    \label{tab:independent}
\end{table*}
Table \ref{tab:independent} shows results corresponding to testing on \heldb dataset and training on \ha. Note that \heldb is more than three times larger than the training data and contains new nodes that have not been seen in \ha.
\epmda, \dbmda, \nimgcn, and \nemii rely on the known miRNA-disease associations to calculate similarities or learn structural embedding for miRNA and disease. Therefore, they cannot be compared in the current inductive setting with new miRNAs or new diseases. 

For such large testing set with many new nodes, \approach still claims its superior performance. With resepct to \ap score, \approach shows an improvement of $\sim$42.4\% over \dimig.

This dataset further supports the importance of the chosen architecture and the effectiveness of multi-sources data integration. 

\subsection{Results on \noveld and \novelmi datasets}

Table \ref{tab:independent} shows results with \noveld and \novelmi test sets. 
\epmda, \dbmda, \nimgcn, and \nemii cannot be used for inductive setting with new nodes. 

 \approach again outperforms its competitor, the \dimig model.
 The gain is more significant on \novelmi datasets with a gain of $\sim48.5\%$ in \ap score.
 On \noveld test set, \approach is $\sim$35.3\% better in \ap score than \dimig.
 
\approach and \dimig share some design similarities as (i) both use GCNs to learn the node representation from the input graph (ii) both incorporate information from the disease ontology, \mi-PCG, disease-PCG and PCG-PCG interaction information. However, there are major differences among the two approaches which dictate the differences in their performance. Firstly, \approach use multiple GCNs to learn the representations for miRNA, disease, and PCG separately from different input graphs while \dimig has a single GCN layer to learn the representation for \mi and PCG from a network build up from PCG-PCG and \mi-PCG interactions while ignoring the heterogeneous character of the network. Secondly,\dimig formulates the \midi association prediction as a multi-label node classification problem and \dimig does not use known \midi associations to train the model; while \approach is a \textit{supervised} multitask model that formulate the \midi association prediction as a binary classification problem. 

Finally, a major difference between the two methods is that \dimig can only work for \mi and disease for which there exist at least one known association with any PCG. \approach does not suffer from such a problem and can work for any set \mi and diseases.

\subsection{Ablation study}
We conduct ablation study to analyze the contribution of additional tasks. The single task baseline employs similar architecture as that of \approach but without the miRNA-PCG and disease-PCG interaction confidence score prediction side tasks. In other words, it also learns miRNA and disease representation from the miRNA family and the disease ontology networks, respectively. However, they only have one classifier layer for the \midi association prediction task, instead of one classifier and two regressors as that of \approach.

Table~\ref{tab:ablation} presents the results for \approach and its single-task variant on both 5-FoldCV and independent testing set up. \approach consistently out-perform its single-task baseline in all datasets. The difference is more significant on the large independent testing sets, especially in \helda and \novelmi. These improvements account for the contribution of the two added side tasks. Since PCGs are the most important link between miRNAs and their associated diseases~\cite{mork2014protein}, miRNA-PCG and disease-PCG prediction tasks also bring additional insights into the \midi association prediction problem.

\begin{table*}[!ht]
    \centering
     \caption{Ablation study results for both 5-FoldCV and independent testing set up.}
  \begin{tabular}{l|c|c|c|c}
    Method & \multicolumn{2}{c}{\textsc{\approach}} & \multicolumn{2}{c}{\textsc{Single task \approach}}\\
    & AUC & AP & AUC & AP\\
    \toprule
    \ha & \textbf{0.833} $\pm$ 0.012 & \textbf{0.832} $\pm$ .015 & 0.827 $\pm$ 0.009 & 0.826 $\pm$ 0.011 \\
\hb & \textbf{0.915} $\pm$ 0.004 & \textbf{0.910} $\pm$ 0.006 & 0.911 $\pm$ 0.004 & 0.906 $\pm$ 0.006 \\
\helda & \textbf{0.705} $\pm$ 0.005 & \textbf{0.691} $\pm$ 0.006 & 0.667 $\pm$ 0.008 & 0.659 $\pm$ 0.007 \\
\novelmi & \textbf{0.726} $\pm$ 0.005 & \textbf{0.714} $\pm$ 0.006 & 0.671 $\pm$ 0.008 & 0.678 $\pm$ 0.006 \\
\noveld & \textbf{0.637} $\pm$ 0.019 & \textbf{ 0.632} $\pm$ 0.025 & 0.603 $\pm$ 0.015 & 0.616 $\pm$ 0.001 \\
\heldb &\textbf{ 0.648} $\pm$ 0.019 & \textbf{0.662} $\pm$ 0.025 & 0.603 $\pm$ 0.015 & 0.653 $\pm$ 0.010 \\
    \end{tabular}
   \label{tab:ablation} 
\end{table*}

\section{Case studies}
\label{resCaseStudy}
To further demonstrate our multi-task model's predictive capability, we evaluate the model for three specific diseases: \bc, \pc, and \dm-type 2. By constructing these case studies, we showcase our model's predictive performance in predicting associations for a specific new disease. To do that, for a specific disease $d$, we select the pairs associated with $d$ from the \hb to use as the testing set. The remaining pairs in \hb are used as the training data. We do negative sampling for both the training and testing set so that the number of positive and negative samples in both training and testing sets are equal. For the test set, the negative pairs are generated only corresponding to the disease $d$.
The dataset statistics for our case studies are presented in table~\ref{tab:stat} where we use the disease names to denotes our generated dataset for that particular disease case study. 

\begin{table}[!htb]
    \centering
    \caption{Our case studies' datasets' statistics, where  $n^+$ and  $n^-$ refer to the number of positive and negative associations/links, respectively.}
    \begin{tabular}{l|c|c|c|c}
         \textsc{Disease} & \multicolumn{2}{c|}{\textsc{Train Samples}} & \multicolumn{2}{c}{\textsc{Train Samples}} \\
         & $n^+$ & $n^-$ & $n^+$ & $n^-$\\
         \hline
         \bc &  8423 & 8423 & 324 & 324 \\
         \pc & 8578 & 8578 & 169 & 169 \\
         \dm & 8640 & 8640 & 107 & 107 \\
    \end{tabular}
    \label{tab:stat}
\end{table}
\begin{figure}[ht!]
    \centering
    \includegraphics[width=0.8\linewidth]{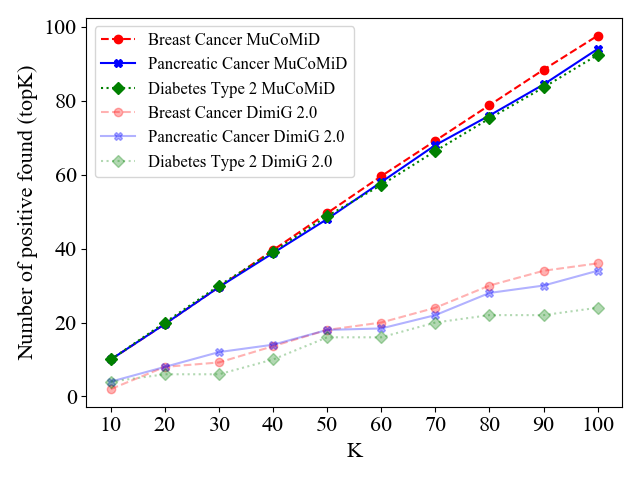}
    \caption{topK interactions for case studies. Mean number of interactions found over 5 independent runs of the models are reported.}
    \label{fig:topk}
\end{figure}
\begin{table*}[!ht]
    \centering
    \caption{MicroRNAs predicted corresponding to top 50 highest prediction scores produced by \approach (average after 5 runs) for the three case studies.}
    \label{tab:top10Result}
    \begin{tabular}{c|l|c|l|c|l|c}
        No. &  \multicolumn{2}{c|}{\bc} &  \multicolumn{2}{c|}{\pc} &  \multicolumn{2}{c}{\dm-type 2} \\
         \hline
         1. & hsa-mir-21 & 0.955 & hsa-mir-21 & 0.956 & hsa-mir-21 & 0.958 \\
2. & hsa-mir-155 & 0.943 & hsa-mir-146a & 0.941 & hsa-mir-155 & 0.951 \\
3. & hsa-mir-146a & 0.938 & hsa-mir-155 & 0.936 & hsa-mir-146a & 0.947 \\
4. & hsa-mir-145 & 0.931 & hsa-mir-146b & 0.933 & hsa-mir-29b & 0.938 \\
5. & hsa-mir-29b & 0.925 & hsa-mir-29b & 0.912 & hsa-mir-145 & 0.935 \\
6. & hsa-mir-146b & 0.921 & hsa-mir-126 & 0.911 & hsa-mir-126 & 0.918 \\
7. & hsa-mir-126 & 0.915 & hsa-mir-145 & 0.907 & hsa-mir-34a & 0.911 \\
8. & hsa-mir-34b & 0.89 & hsa-mir-210 & 0.872 & hsa-mir-142 & 0.903 \\
9. & hsa-mir-150 & 0.889 & hsa-mir-150 & 0.871 & hsa-mir-221 & 0.902 \\
10. & hsa-mir-34a & 0.887 & hsa-mir-34a & 0.87 & hsa-mir-150 & 0.897 \\
11. & hsa-mir-142 & 0.885 & hsa-mir-142 & 0.867 & hsa-mir-34c & 0.896 \\
12. & hsa-mir-34c & 0.882 & hsa-mir-223 & 0.862 & hsa-mir-223 & 0.889 \\
13. & hsa-mir-210 & 0.881 & hsa-mir-34c & 0.857 & hsa-mir-122 & 0.886 \\
14. & hsa-mir-223 & 0.87 & hsa-mir-34b & 0.856 & hsa-mir-222 & 0.869 \\
15. & hsa-mir-221 & 0.868 & hsa-mir-122 & 0.835 & hsa-mir-26a & 0.864 \\
16. & hsa-mir-122 & 0.862 & hsa-mir-206 & 0.83 & hsa-mir-205 & 0.862 \\
17. & hsa-mir-205 & 0.859 & hsa-mir-221 & 0.813 & hsa-mir-29a & 0.853 \\
18. & hsa-mir-31 & 0.857 & hsa-mir-31 & 0.811 & hsa-mir-375 & 0.85 \\
19. & hsa-mir-206 & 0.835 & hsa-mir-143 & 0.809 & hsa-mir-144 & 0.848 \\
20. & hsa-mir-144 & 0.832 & hsa-mir-205 & 0.805 & hsa-mir-143 & 0.846 \\
21. & hsa-mir-29a & 0.832 & hsa-mir-375 & 0.802 & hsa-mir-206 & 0.844 \\
22. & hsa-mir-143 & 0.83 & hsa-mir-222 & 0.795 & hsa-mir-27b & 0.831 \\
23. & hsa-mir-124 & 0.828 & hsa-mir-26a & 0.794 & hsa-mir-24 & 0.83 \\
24. & hsa-mir-26a & 0.827 & hsa-mir-183 & 0.785 & hsa-mir-192 & 0.826 \\
25. & hsa-mir-222 & 0.822 & hsa-mir-181a & 0.785 & hsa-mir-183 & 0.824 \\
26. & hsa-mir-375 & 0.821 & hsa-mir-24 & 0.782 & hsa-mir-27a & 0.819 \\
27. & hsa-mir-183 & 0.818 & hsa-mir-29a & 0.776 & hsa-mir-19b & 0.816 \\
28. & hsa-mir-24 & 0.811 & hsa-mir-22 & 0.748 & hsa-mir-1 & 0.813 \\
29. & hsa-mir-132 & 0.807 & hsa-mir-19a & 0.743 & hsa-mir-214 & 0.808 \\
30. & hsa-mir-1 & 0.797 & hsa-mir-20a & 0.742 & hsa-mir-9 & 0.807 \\
31. & hsa-mir-192 & 0.796 & hsa-mir-204 & 0.74 & hsa-mir-19a & 0.802 \\
32. & hsa-mir-19b & 0.795 & hsa-mir-29b-1 & 0.738 & hsa-mir-17 & 0.798 \\
33. & hsa-mir-29b-1 & 0.793 & hsa-mir-106b & 0.735 & hsa-mir-20a & 0.798 \\
34. & hsa-mir-29c & 0.792 & hsa-mir-29c & 0.735 & hsa-mir-92a & 0.797 \\
35. & hsa-mir-181a & 0.79 & hsa-mir-17 & 0.734 & hsa-mir-22 & 0.797 \\
36. & hsa-mir-27b & 0.789 & hsa-mir-106a & 0.734 & hsa-mir-93 & 0.795 \\
37. & hsa-mir-196a & 0.785 & hsa-mir-18a & 0.732 & hsa-mir-182 & 0.792 \\
38. & hsa-mir-182 & 0.78 & hsa-mir-132 & 0.727 & hsa-mir-18a & 0.792 \\
39. & hsa-mir-19a & 0.78 & hsa-mir-182 & 0.722 & hsa-mir-132 & 0.791 \\
40. & hsa-mir-22 & 0.779 & hsa-mir-18b & 0.717 & hsa-mir-18b & 0.788 \\
41. & hsa-mir-17 & 0.777 & hsa-mir-19b-1 & 0.705 & hsa-mir-20b & 0.785 \\
42. & hsa-mir-20a & 0.777 & hsa-mir-27b & 0.698 & hsa-mir-23b & 0.785 \\
43. & hsa-mir-9 & 0.776 & hsa-mir-16 & 0.688 & hsa-mir-486 & 0.782 \\
44. & hsa-mir-92a & 0.774 & hsa-mir-15a & 0.688 & hsa-mir-15a & 0.775 \\
45. & hsa-mir-15a & 0.773 & hsa-mir-92a & 0.686 & hsa-mir-195 & 0.771 \\
46. & hsa-mir-214 & 0.772 & hsa-mir-342 & 0.683 & hsa-mir-204 & 0.767 \\
47. & hsa-mir-106b & 0.771 & hsa-mir-214 & 0.682 & hsa-mir-320a & 0.766 \\
48. & hsa-mir-29b-2 & 0.769 & hsa-mir-133b & 0.68 & hsa-mir-23a & 0.757 \\
49. & hsa-mir-16 & 0.769 & hsa-mir-9 & 0.678 & hsa-mir-15b & 0.749 \\
50. & hsa-mir-93 & 0.768 & hsa-mir-96 & 0.673 & hsa-mir-133b & 0.749 \\

    \end{tabular}
    
\end{table*}
Figure~\ref{fig:topk} presents the topK evaluation results while table~\ref{tab:top10Result} shows the top 50 predictions of the associated miRNAs for the three diseases. For each case study, the known association for that particular disease is completely hidden from the model training process. The statistics of our training and testing data can be found in Section~\ref{tab:stat}. 

Looking at Figure~\ref{fig:topk}, though the case studies' diseases are completely new, our model still attains near-perfect prediction for the top 40 predictions. Compared with \dimig for the top 50 predictions, our model has a gain of at least $166\%$.

For \bc, the model acquires $\sim 97.6\%$ accuracy for the top 100 predictions. For \dm-type 2, the number of known positive associations is only 107, but out of the top 100 predictions, we can correctly recognize $92.4$ associations (this number is the average of the number of found interactions over five runs). These results again claim the effectiveness of our model in predicting potential associations for new diseases.

\section{Conclusion}
\label{secConclusion}
We propose a multi-task graph convolutional learning framework, \approach for the problem of predicting \mi-disease associations.
Our end-to-end learning approach allows automatic feature extraction while incorporating knowledge from five heterogeneous biological information sources. Incorporating many sources of information helps compensate for the lack of information in any single source and, at the same time, enables the model to generate predictions for any new miRNA or disease. Unlike previous works, our model can be employed in both transductive and inductive settings. To test the generalization power of models, we construct and test on both existing benchmarked set up and on larger independent test sets. Large-scale experiments in several testing scenarios highlight the superiority of our approach. An ablation study is added to highlight the side tasks' contribution. We release all the code and data used in this study for reproducibility and future research at~\href{https://git.l3s.uni-hannover.de/dong/cmtt}{\url{https://git.l3s.uni-hannover.de/dong/cmtt}}.

We believe that our design principles will be of independent interest for other biomedical applications where data scarcity is a major challenge. In particular, the use of multi-task learning to integrate information from heterogeneous information sources to overcome the problems of data scarcity and unreliability of one single data type is a unique perspective and has not been studied for computational problems in biomedicine.

\ifCLASSOPTIONcompsoc
  \section*{Acknowledgments}
\else
  
  \section*{Acknowledgment}
\fi
The work is partially supported by the PRESENt project funded by Volkswagen Stiftung and the State Government of Lower Saxony (grant no. 11-76251-99-3/19 (ZN3434)), the Federal Ministry of Education and Research (BMBF), Germany under the LeibnizKILabor project (grant no. 01DD20003), and the strategic funds of the Translational Alliance in Lower Saxony.

\bibliographystyle{IEEEtran}
\bibliography{acmart}
\pagebreak
\begin{IEEEbiography}[{\includegraphics[width=1in,height=1.25in,clip,keepaspectratio]{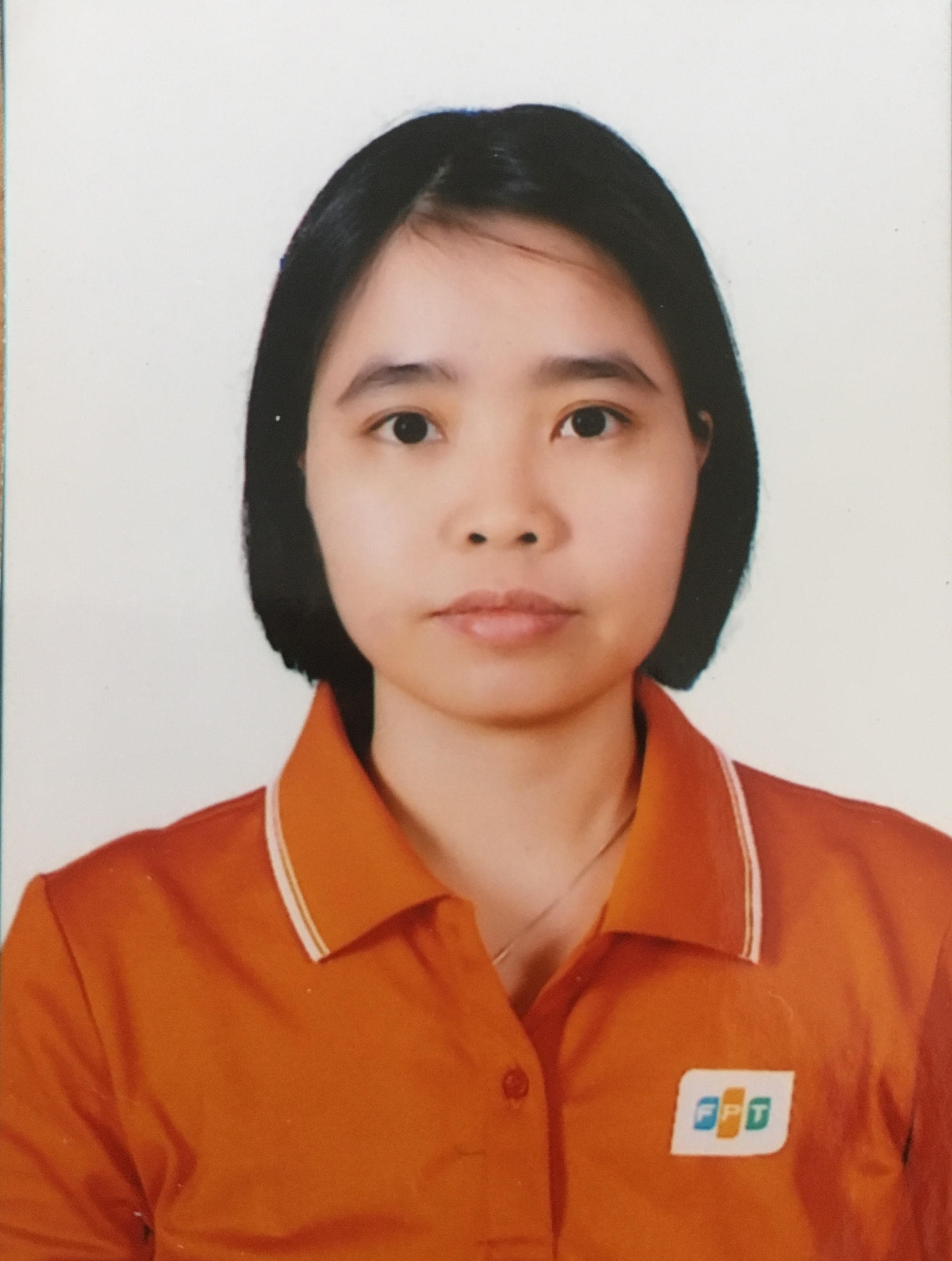}}]{Ngan Dong} is PhD student at L3S Research Center, Leibniz University of Hannover, Germany. She works as a research assistant in the~\href{http://www.translationsallianz.de/train-platforms/train-projects/present/?L=1}{PRESENt} project\footnote{http://www.translationsallianz.de/train-platforms/train-projects/present/?L=1}, which aims at integrating clinical, biological, and big data research to advance our understanding of norovirus gastroenteritis. Her current research focuses on network analysis, feature selection, graph-based representation learning, multi-task models, and data integration.
\end{IEEEbiography}
\begin{IEEEbiography}[{\includegraphics[width=1in,height=1.25in,clip,keepaspectratio]{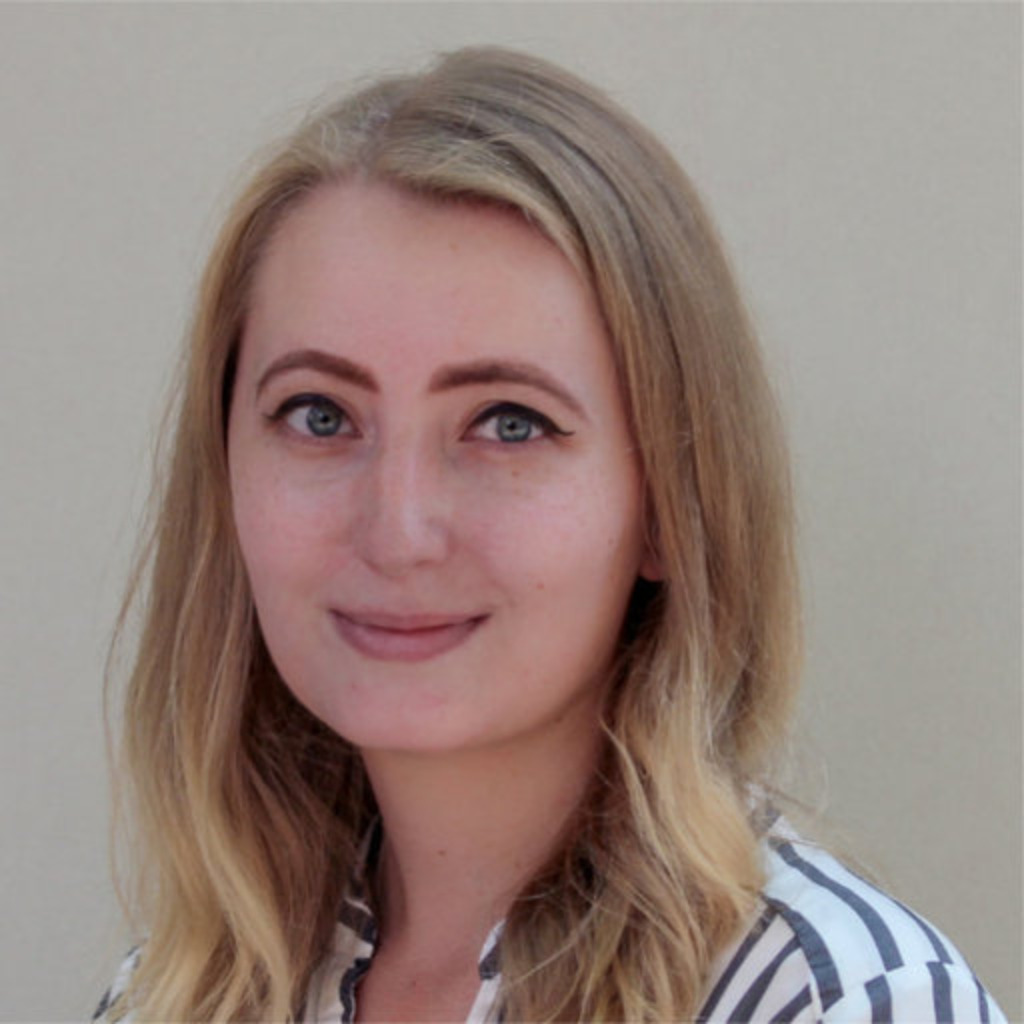}}]{ Stefanie Mücke} received her B. Sc. and M. Sc. in Biology at the Martin-Luther-Universität Halle/Wittenberg (Germany) in 2013 and 2016, respectively. In 2020, she earned her Ph.D. in Biotechnology at the Gottfried Wilhelm Leibniz Universität Hannover (Germany), working on host-pathogen interactions, omics analyses, genetic engineering, and biotechnology. Since 2020, she has been a scientific officer for the Translational Alliance in Lower Saxony, a network of 10 research institutions in Hannover and Braunschweig (Germany). She supports cooperation projects between these institutions in the capacity of a project manager and scientific advisor.
\end{IEEEbiography}
 \begin{IEEEbiography}[{\includegraphics[width=1in,height=1.25in,clip,keepaspectratio]{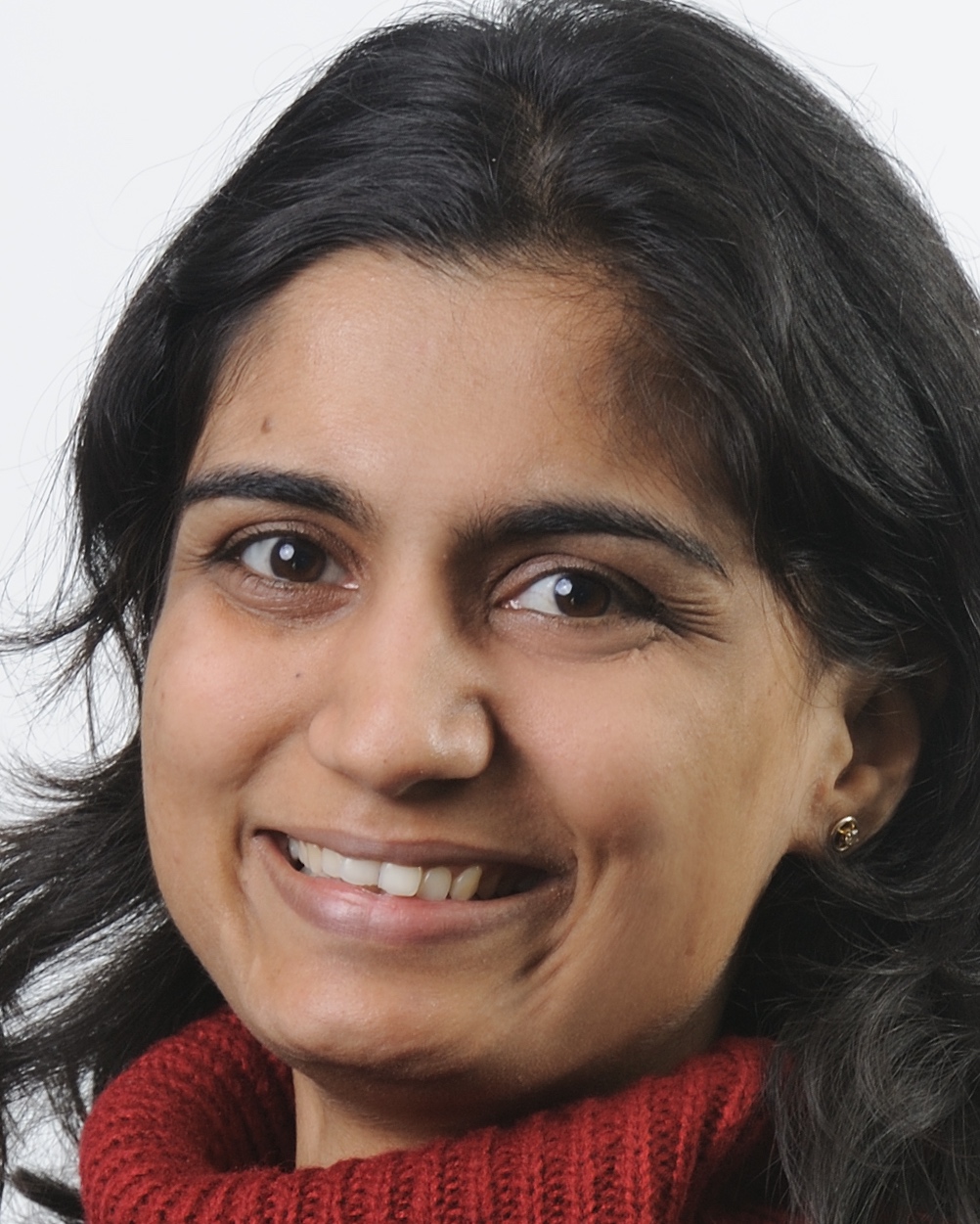}}]{Megha Khosla} received her PhD degree in Theoretical Computer Science from Max Planck Institute of Informatics and Saarland University, Saarbruecken, Germany. Currently she is a senior researcher at L3S Research center. Her main research focus is on effective, interpretable and privacy-preserving learning on graphs with applications to personalized medicine.
\end{IEEEbiography}
\vfill
\end{document}